# Correlation between surface rumpling and structural phase transformation of $SrTiO_3$[A]

S. Singh[1,2], Te-Yu Chien(簡德宇)[3], J. R. Guest[4] and M. R. Fitzsimmons[1]

[1]*Los Alamos National Laboratory, Los Alamos, NM 87545, USA*

[2]*Solid State Physics Division, Bhabha Atomic Research Center, Mumbai 400085 India*

[3]*Advanced Photon Source, Argonne National Laboratory, Argonne, IL 60439 USA*

[4]*Center for Nanoscale Materials, Argonne National Laboratory, Argonne, IL 60439 USA*

**Abstract**

We present x-ray reflectivity, x-ray diffraction and atomic force microscopy measurements of single crystal $SrTiO_3$ taken as a function of temperature. We found a rumpling transformation of the $SrTiO_3$ surface after cooling the sample below ~105 K. The rumpling transformation is correlated with the cubic to tetragonal phase transformation that occurs at the same temperature. The rumpling transformation is reversible.

**Introduction**

Strontium titanate, $SrTiO_3$, is a model example of a perovskite-structured oxide [1] with remarkable multi-functional properties. In stoichiometric form, $SrTiO_3$ is an optically transparent insulator with a high dielectric constant and therefore is suitable as an insulating layer in high $T_c$ multilayer structures for fundamental research and device applications [2-4]. Examples include: high-temperature superconductors, colossal magnetoresistive materials, ferroelectrics, and heterostructures containing two-dimensional electron gases [5-8]. $SrTiO_3$ has many attractive features for epitaxial thin film growth including: low cost compared to other perovskite single crystals, a lattice parameter that is reasonably well-matched to many perovskite films, relative absence of twins, and a non-polar structure that is conducive to smooth surfaces [9-16]. Thus, $SrTiO_3$ has been extensively studied [9-23].

$SrTiO_3$ undergoes a structural phase transition from cubic (a = 3.905Å) to tetragonal (c/a = 1.00056) phase at ~105 K [24-26]. The phase transition involves the rotation of $TiO_6$ octahedra and has been featured as a classic example of a soft mode phase transition [24]. There have been conflicting reports [17-21, 27-29] that a surface phase transition may occur at significantly higher temperature ~ 150 K and a second bulk phase transformations below ~70 K [25, 26]. The "surface" structures of $SrTiO_3$ reported in ref. [19, 20] were measured using X-ray diffraction (XRD) with a penetration depth of 0.4 to 57 μm. Some reports [17, 21] have suggested that structure of the $SrTiO_3$ surface may be affected by the quality of the $SrTiO_3$ crystal. Here, we used two probes of surface structure, X-ray reflectivity (XRR) [30-32] and atomic force microscopy (AFM) [33] to characterize the surface structure of $SrTiO_3$ as a function of temperature. The observations are correlated with those of the bulk structure inferred from XRD.

A disadvantage of $SrTiO_3$ is that the cubic to tetragonal structural phase transformation may induce strain or microcracks in a film grown on a $SrTiO_3$ substrate. Strain or microcracks may influence strain-sensitive properties, e.g., magnetoresistance, multiferroicity, ferroelectricity, etc. In addition to the change of atomic structure, we found concomitant severe rumpling of the $SrTiO_3$ surface. The rumpling of the substrate may play an even larger role in affecting the properties of films deposited onto the $SrTiO_3$ than that caused by the change of atomic structure.

**Experimental**

Single crystal $SrTiO_3$ (001) measuring $10 \times 10 \times 1$ $mm^3$ grown by the Verneuil technique [34] was obtained from CrysTec GmbH (Berlin). Substrates such as these have been extensively used by the research community to grow epitaxial complex oxide films. The (001) surface was polished by the vendor to be suitable for epitaxial film growth. The $SrTiO_3$ crystal was ultrasonically degreased for 10 min in acetone and isopropanol. The substrate was sonicated in deionized water (18 MΩ/cm) for 15 min and subsequently, loaded into a tube furnace with flowing $O_2$ and annealed at 950 °C for 5 h. X-ray scattering measurements were performed as a function of temperature using a closed cycle helium cryostat. Temperature was varied between room temperature and 10 K with an accuracy of better than 0.1 K. X-ray scattering measurements were carried out using Cu $K_\alpha$ radiation at Los Alamos Neutron Science Center (LANSCE). X-ray measurements were taken with the projection of the incident x-ray wave vector on the sample's surface parallel to [001], [110] and [010] $SrTiO_3$. We show only data taken for the first orientation, since we observed no significant change in any of our findings depending upon the orientation of the substrate about its surface normal.

XRR is a nondestructive technique from which the depth dependent structure of the sample with nanometer resolution averaged over the lateral dimensions of the entire sample (typically 100 mm$^2$) can be inferred [30-32]. XRR involves measurement of the x-ray radiation reflected from a sample (Fig. 1) as a function of wave vector transfer $Q$ (i.e., the difference between the outgoing and incoming wave vectors). The most intensely reflected beam (thick black arrow Fig. 1a) corresponds to the specular reflectivity where the angle of reflection from the surface $\theta_f$ and the angle of incidence $\theta_i$ are equal. Note that rumpling or faceting of the sample's surface over large lateral dimensions will change $\theta_i$ and thus change $\theta_f$ equally for the specular reflection. In addition, off-specularly scattered (diffuse) radiation (i.e., that producing a range of $\theta_f$ for a single $\theta_i$ can be observed (thin blue arrows, Fig. 1a) when the height fluctuations of the surface are correlated along the lateral dimensions of the surface [30]. We used a linear position sensitive detector (PSD) to simultaneously measure the specular and off-specular x-ray reflectivity over a large range of wave vector transfer parallel, $Q_x$ [= $\frac{2\pi}{\lambda}\left(cos(\theta_i) - cos(\theta_f)\right)$, where $\lambda$ is wavelength of x-ray] and perpendicular, $Q_z$ [= $\frac{2\pi}{\lambda}\left(sin(\theta_i) + sin(\theta_f)\right)$], to the sample's surface (Fig. 1 (b)).

Before changing temperature, radial and transverse scans in reciprocal space were also recorded with XRD. The widths of Bragg reflections along the radial (longitudinal) direction in reciprocal space provide information about microstrain and grain size in the sample corresponding to atomic length scales [35]. The transverse scan provides information about changes to the crystalline quality, e.g., mosaic spread, of the sample.

The temperature dependent topography of the sample surface was measured using a commercial variable temperature UHV atomic force microscope (VT-UHV-AFM/STM) system in contact mode at different temperatures.

## Results

*Temperature dependence of the broadening and splitting of the specular reflectivity*

Figs. 2(a)-(d) show XRR ($Q_x$ - $Q_z$ scattering map at small angle) from the (001) surface of the SrTiO$_3$ crystal at temperatures of 290, 105, 77 and 10 K, respectively (though we measured the XRR at many temperatures). The x-ray intensity as a function of $Q_z$ at $Q_x = 0$ (vertical dash-dash line in Fig. 2(a)) corresponds to the specular XRR. The x-ray intensity as a function of $Q_x$ at fixed $Q_z$ (horizontal dash-dash line in Fig. 2(a)) corresponds to off-specular XRR [30-31]. Figs. 2(a)-(d) show that while cooling the sample, the specular reflection (centered about $Q_x = 0$) broadens along $Q_x$, and splits into multiple reflections indicating significantly increased in-plane inhomogeneity. The temperature dependence of the XRR data for $Q_z \sim 0.1138$ Å$^{-1}$ is plotted vs. $Q_x$ in Fig. 2(e). Above 150 K the specular reflection was sharp. A small broadening of the specular reflection about $Q_x = 0$ was first observed at a temperature of 150 K and with further cooling, the specular reflection broadened dramatically. While warming the sample from low temperatures, the reverse behavior was observed, i.e., the reflection sharpened at high temperature. Thus, the surface transition from smooth to rumpled surface was reversible with temperature. The broadening (FWHM) of specular XRR for a fixed angle of incidence was obtained from the entire $Q_x$ profile [thus, for cases where multiple sharp peaks (specular reflections) were observed the outermost peaks determined the FWHM]. The FWHM of the specularly reflected x-ray beam at an angle of incidence ($\theta_i = 0.8°$) are shown in Fig. 3(a) while cooling and warming. The FWHM of specular XRR during cooling and warming shows similar temperature dependence.

*Temperature and large $Q_x$ dependence of the off specular (diffuse) scattering*

In contrast to the behavior of the XRR for small $Q_x$, for large $Q_x$, e.g., $Q_x > \pm 2.0 \times 10^{-4}$ Å$^{-1}$, the $Q_x$ dependence of the XRR is affected very little with temperature. Using a self-affine fractal surface model [30] to represent the diffuse scattering for large $Q_x$ (solid line in Fig. 2(e)), we obtained topographic parameters, $\sigma$ (root mean square roughness amplitude), $h$ (Hurst Parameter) and $\xi$ (in-plane correlation length) of 0.4±0.1 nm, 0.45±0.05 and 0.8±0.05 μm, respectively, for the (001) surface of SrTiO$_3$. A Hurst parameter of 0.5 corresponds to a Gaussian distributed height-height correlation function across the sample surface [30]. The parameters $\sigma$, $h$, and $\xi$ are ones that describe atomic-scale roughness that is correlated over the lateral dimensions of the sample at submicron length scales (i.e., length scales of order $\xi$).

*Temperature and $Q_z$ dependence of the specular reflectivity*

The specular XRR as a function of $Q_z$, ($R(Q_z)$), is qualitatively related to the Fourier transform of the scattering length density (SLD) depth profile $\rho(z)$ [30-31] averaged over the sample's lateral dimensions. $\rho(z)$ is proportional to the electron density [30-31]. Surface roughness (height–height fluctuation of the surface, $\sigma'$ over all lateral length scales) is another parameter which influences $R(Q_z)$ in a manner similar to a thermal or static Debye-Waller factor [35]—the intensity is exponentially damped by a factor that varies as $-\sigma'^2 Q_z^2/2$. By definition specular reflectivity means $Q_x = 0$, thus, the decay of $R$ with $Q_z$ is related to the roughness of the surface over all lateral length scales [30-31, 36]. $R(Q_z)$ vs. temperature measured during cooling is plotted in Fig. 2(f). The inset of Fig 2(f) shows the electron SLD profile that gave the best fit to the specular reflectivity taken at 290 K. The fit yielded a value of electron SLD, which corresponds to the bulk value of single crystal of SrTiO$_3$, and a value of the surface roughness equal to $\sigma'=0.4\pm0.1$ nm. We observed little change in the $Q_z$-dependence of the specular XRR while cooling the sample until ~77 K, suggesting the surface roughness at atomic length scales is not changing above 77 K (while the surface rumpling at micron+ length scales does change). At low temperatures (below 77 K), the decay of the reflectivity with $Q_z$ increases suggesting an increase of surface roughness, which nearly doubles from ~0.4 nm to ~0.7 nm.

*AFM and XRD measurements*

We used AFM to obtain images of atomic-scale roughness of our $SrTiO_3$ sample as a function of temperature (Fig. 4). Figs. 4 (a-d) show topographical AFM images with a scan size of $1 \times 1 \mu m^2$ at 300, 210, 100 and 61 K, respectively. Figs. 4 (e-h) show the height information along the line-cut of the AFM images [Figs. 4 (a-d)] at 300, 210, 100 and 61 K, respectively. The AFM measurements were performed in the following order: 300, 61, 100 and 210 K. The topography exhibits terraces with widths of about 100 nm and steps of ~0.5 nm. At 210 and 300 K we observed well defined terraces with small fluctuation of height (Fig. 4(d)) on their surfaces. At low temperatures the topography was noticeably different. The fluctuation in the heights (roughness) of the terraces (Fig. 4 (f and g)) increased, and the atomic-scale variation on the surfaces of the terraces became more irregular (Fig. 4(c) and (d)). All images show a systematic variation of the height over length scales much larger than the scanned dimension of 1 micron.

In conjunction with the XRR study, we took XRD measurements. Specifically, after collecting XRR data, we collected XRD data, then changed temperature and repeated the data collection process. Measurements of the longitudinal width (a measure of microstrain and/or grain size) and rocking curve width (a measure of crystal quality) of the (004) $SrTiO_3$ Bragg reflection for (obtained with XRD) are shown in Figs. 3(b and c), respectively. The data are plotted for cooling and warming cycles.

**Discussion**

A comparison of the FWHM's for the specular reflection, radial and rocking curves (Figs. 3(a) (b) and (c)) show sharp increases in the FWHM's beginning around 105 K. The correlation between the rumpling (as measured by the broadening of the specular reflection Figs. 3(a) and change of atomic structure (as observed with XRD Figs. 3(b) and (c)) imply the two effects are correlated. We conclude the phase transition of the bulk atomic structure [24-28] induces rumpling of the [001] $SrTiO_3$ surface at 105 K.

The broadening and splitting of the reflected x-ray intensity at low temperatures (mostly below 105 K) in the $Q_x$ - $Q_z$ scattering maps (Figs. 2(b) – (d)) are consequences of incoherent superposition of reflection from surfaces with long waviness (surface facets). Facets tilted with respect to each other as depicted in Fig. 5(a) are an example of a rumpled surface that could produce the splitting of the specular reflection we observed. The sizes of these surface facets (waviness) exceed the lateral coherence of the x-ray beam which is of order of a few micrometers [32]. The tilting of facets [splitting of the specular reflection in Fig. 2(e)] increases with decreasing the temperature. The angular spread of the facets is not more than 0.5°. Thus in addition to atomic scale roughness as exemplified by ($\sigma$, $\xi$ and $h$) (Fig. 5(b) and AFM images Fig. 4), we observed changes at and below 105 K to the surface structure occurring over larger lateral length scales, i.e., rumpling (facets) of micron+ size tilted with respect to each other by ~0.5° as represented in Fig. 5(a).

In contrast to the surface rumpling, which occurs over long lateral length scales, the atomic-scale roughness as measured by the decay of the specular XRR, is relatively unaffected by the cubic-tetragonal phase transition of the film bulk (at 105 K). The roughness we found is close to the lattice parameter (~0.4 nm) of $SrTiO_3$ and consistent with the steps in the AFM line profiles (Figs. 4d-f). The atomic scale roughness, $\sigma'$, obtained from specular XRR as a function of temperature while warming (similar in cooling cycle) are shown in Fig. 5 (c).

Since we lack AFM data over micron+ length scales, we cannot measure the standard deviation of the height over the micron+ length scales from the AFM data. Consequently, an estimate of roughness based upon the heights in the AFM data will include sources to the variance from the systematic trend, which is ill-defined as discussed previously, and the atomic-scale roughness. In order to obtain an estimate of the atomic-scale roughness from the AFM data, we fitted the data to a surface defined by a polynomial of degree two. We subtracted the surface from the AFM data, and then calculated histograms of atomic-scale fluctuations (Fig. 5(d)). The *rms* width of the histograms and the error on the standard deviation are shown in Fig. 5(c). The histogram for the 61 K data is noticeably broader than the other histograms taken at higher temperatures. This trend with temperature observed in the AFM data is consistent with the trend seen in $\sigma'$ obtained from XRR. We expect $\sigma'$ obtained from XRR to be larger than the roughness estimate obtained from AFM because all lateral length scales contribute to $\sigma'$.

The width of the specular XRR broadens most profoundly at 105 K. The broadening suggests that the surface becomes rumpled at micron+ lateral length scales at 105 K. Yet, the $Q_z$ decay of the intensity of the specular XRR does not begin until temperatures below 77 K. The onset of the decay suggests that atomic scale roughening of the surface does not begin until 77 K (consistent with the AFM results). The XRR measurements suggest another modification of the

(001) SrTiO$_3$ surface may occur at very low temperatures, and the modification is different from surface rumpling occurring at 105 K.

**Conclusion**

In conclusion, we measured the temperature dependent change of the surface of a single crystal SrTiO$_3$ (001) using low and wide angle x-ray scattering and AFM. Below 105 K the surface rumples (or becomes wavy) over micron+ length scales. The magnitude of the rumpling (tilting of facets) increases with decreasing temperature to a value as large as 0.5°. The rumpling of the surface is correlated with a change of atomic structure as observed by a degradation of the mosaic quality of the crystal substrate (inferred from the rocking curve width) and increased the microstrain and/or decreased the grain size of the crystal (inferred from the radial width of a Bragg reflection). The degradations of surface and bulk crystal quality increase as temperature decreases below 105 K. These observations are not unique to the sample reported here; rather, the vast majority of SrTiO$_3$ substrates observed during the more than decade long Asterix user program at LANSCE [37] exhibit the rumpling behavior. The preponderance of surface rumpling of SrTiO$_3$ may be a source of concern for studies of strain sensitive properties, e.g., ferroelectricity, magnetism, etc, of films grown on SrTiO$_3$.

**Acknowledgements**

This work was supported by the Office of Basic Energy Science, U.S. Department of Energy, BES-DMS funded by the Department of Energy's Office of Basic Energy Science. Los Alamos National Laboratory is operated by Los Alamos National Security LLC under DOE Contract DE-AC52-06NA25396. Use of the Center for Nanoscale Materials was supported by the U. S.

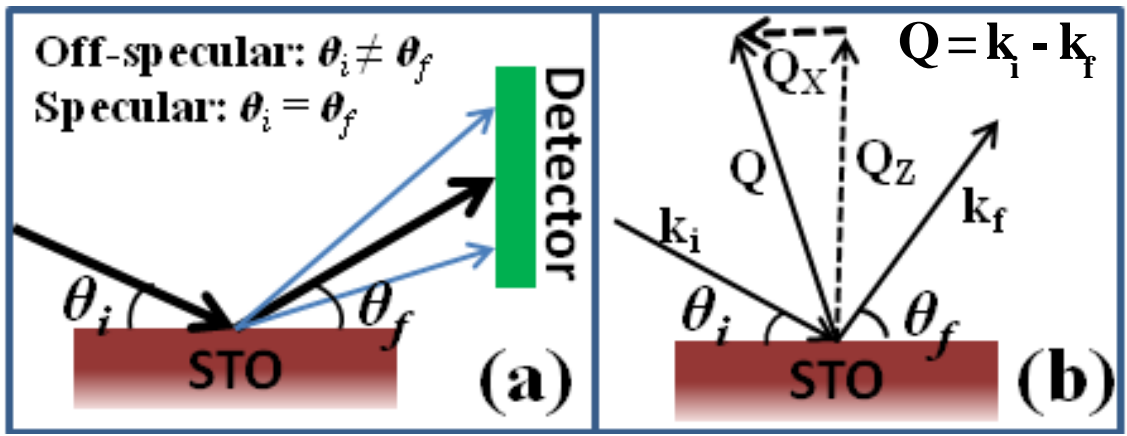

Fig. 1 (a): X-ray scattering geometry using position sensitive detector. (b) Scattering geometry in reciprocal space.

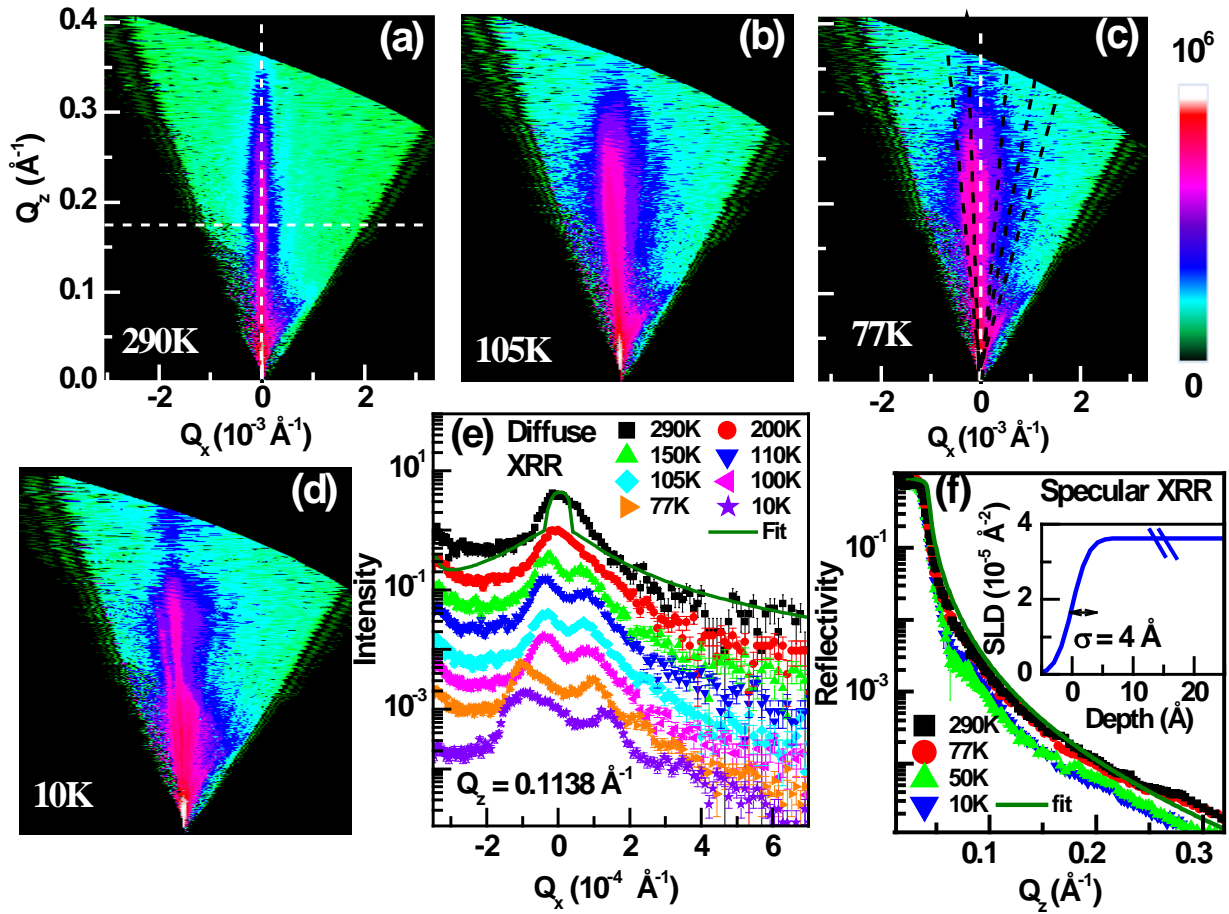

Fig. 2 (a)-(d): Reciprocal space map ($Q_x – Q_z$ map) of x-ray reflectivity at different temperatures on cooling of the SrTiO$_3$ crystal. Dash-dash line (white) along $Q_x = 0$ in Fig (c) shows the specular reflectivity whereas the dash-dash lines (black) show the reflected intensity from large waviness roughness which is giving broadening at specular ridge. Temperature dependent diffuse XRR (e) at $Q_z = 0.1138$ Å$^{-1}$ and specular XRR (f). Diffuse XRR in Fig. (e) are shifted by a factor of ~2.5 for clear visibility. Inset of Fig. (f) shows the electron scattering length density (SLD) depth profile which gave best fit to specular XRR data measured on cooling sample from 290 K to 77 K.

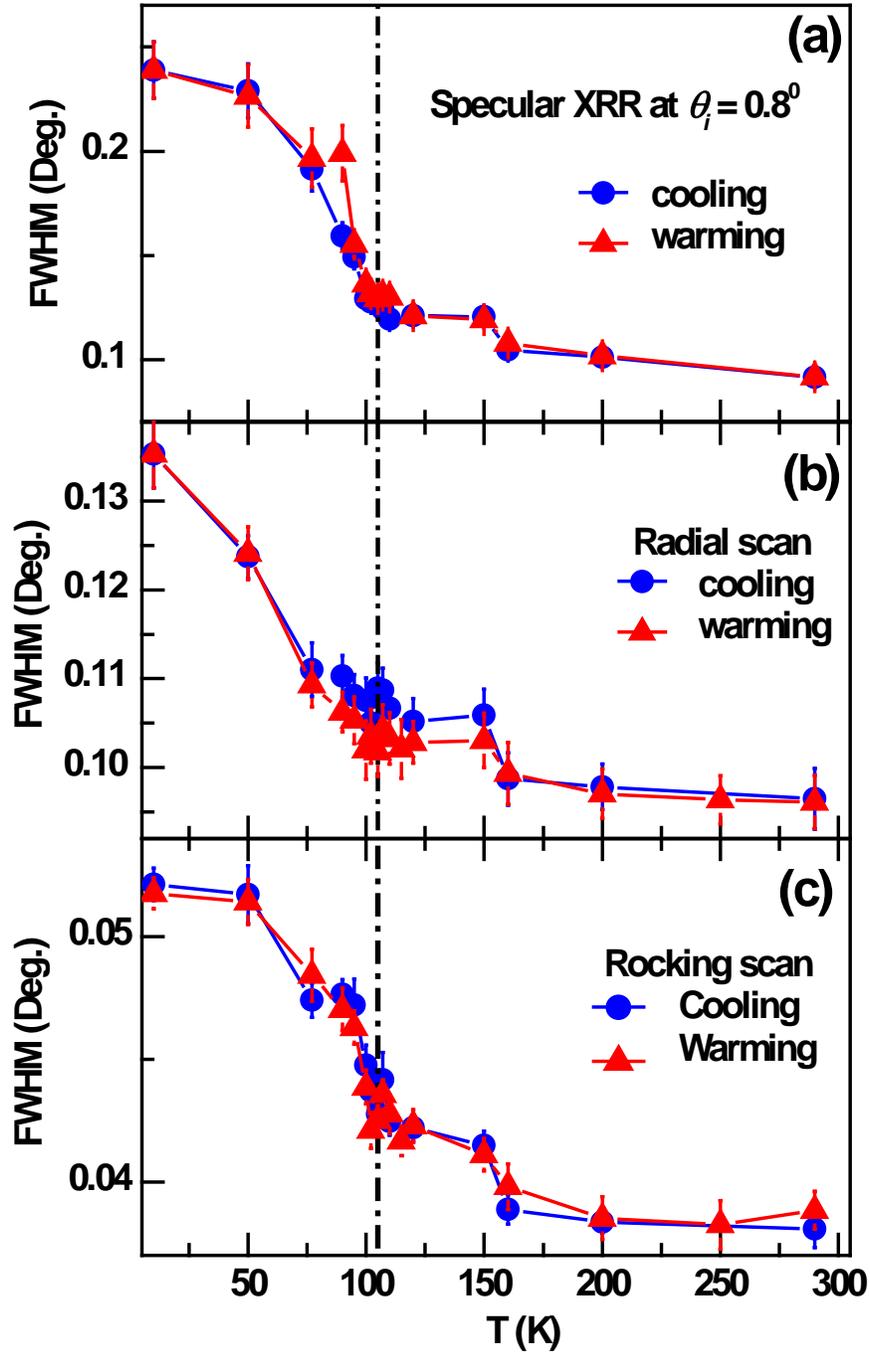

Fig. 3: Temperature dependent variation of Full width at half maxima (FWHM) of specular x-ray reflection (a) at an angle of incidence, $0.8^0$. Temperature dependent variation of FWHM of (004) reflection of $SrTiO_3$ by x-ray diffraction in radial (b) and rocking scan (c).

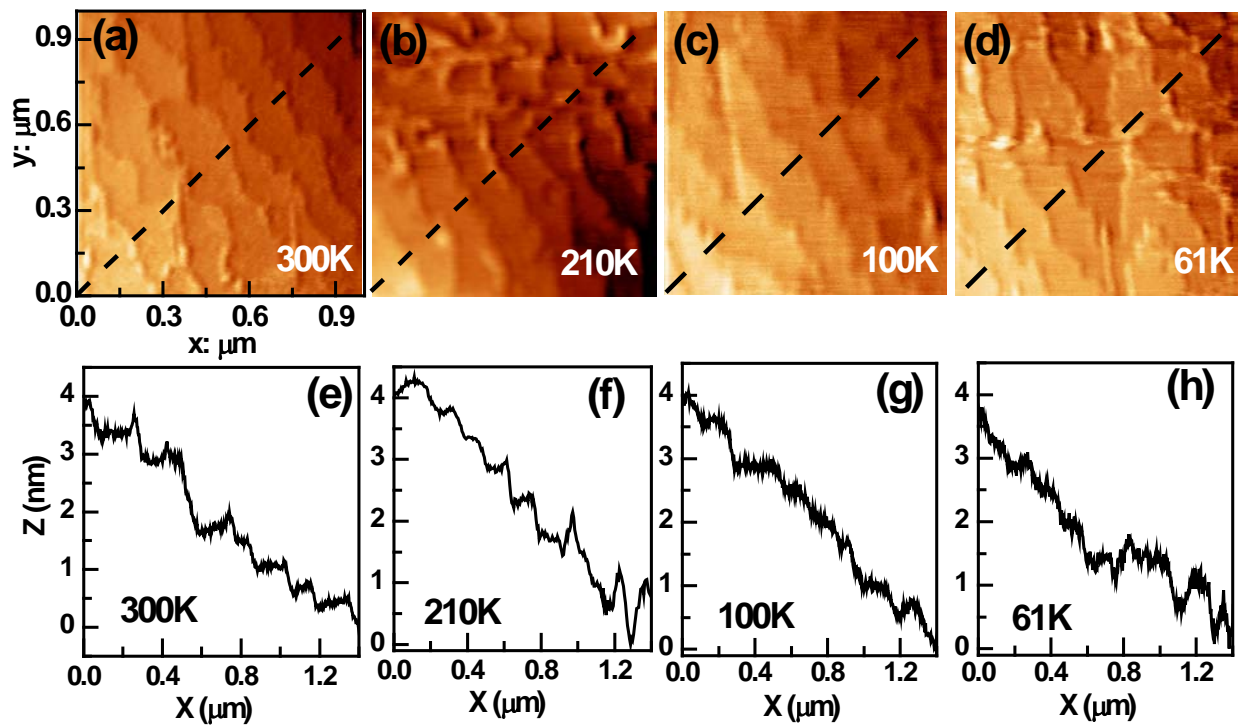

Fig. 4: AFM measurements (1 × 1 µm²) at different temperature of 300 (a), 210 (b), 100 (c) and 61 K (d) from SrTiO$_3$ surface. The AFM measurements were performed in the following order: 300, 61, 100 and 210 K. Fig. (e), (f), (g) and (h) show the height information along a line marked in Figs. 4(a), (b), (c) and (d), respectively. The AFM images (Fig. 4 (a-d)) are represented using derivative enhanced images. However the line profiles (Fig. 4 (e-h)) are taken from raw AFM Data.

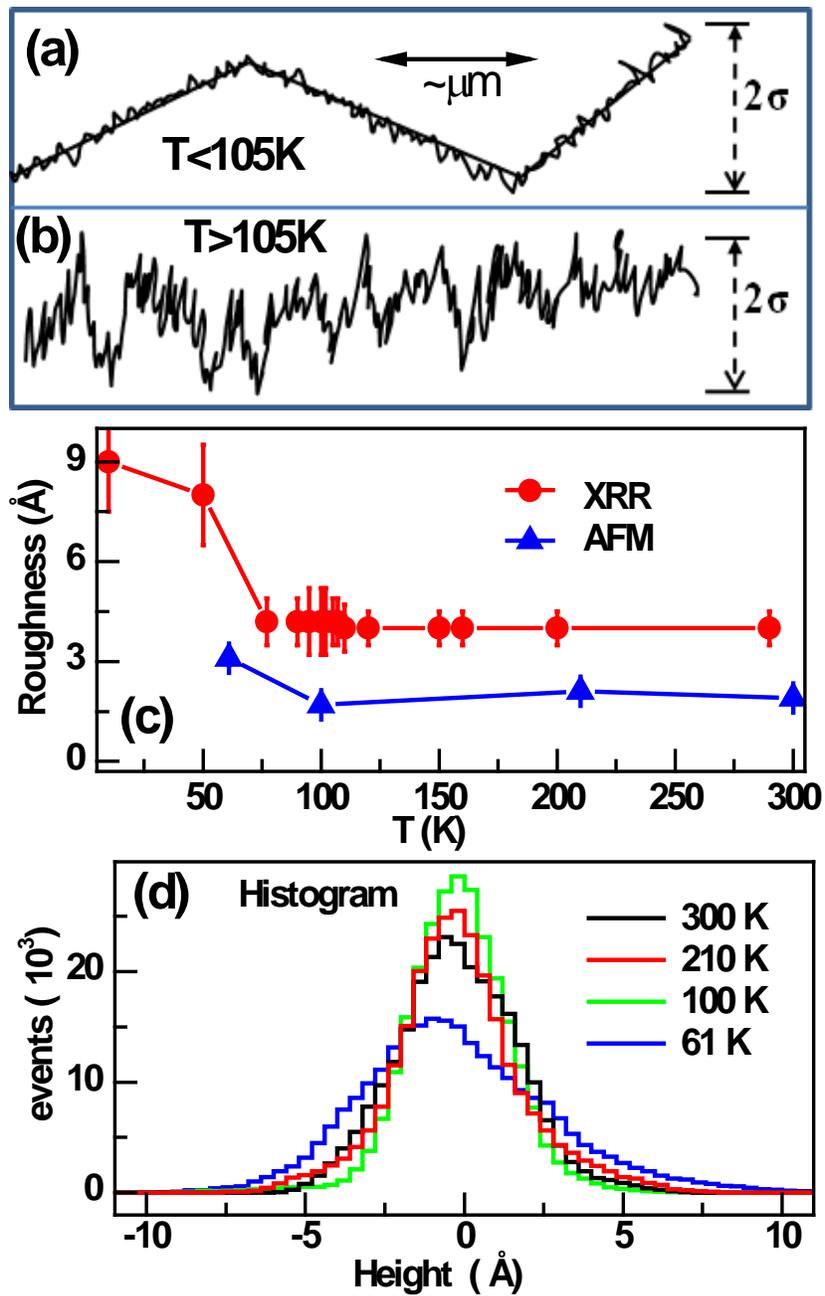

Fig. 5: Representation of surface modification inferred from X-ray scattering measurements below (a) and above (b) 105 K. (c): variation of roughness with temperature measured using XRR (●) and AFM (▲). (d) Histograms showing the fluctuation of surface height after removal of a systematic trend (discussed in text).